\documentclass[%
 preprint,
 amsmath,amssymb,
 aps,
]{revtex4-2}
\usepackage{makecell}
\usepackage{graphicx}
\usepackage{dcolumn}
\usepackage{bm}
\usepackage{hyperref}
\usepackage{float}

\begin{document}

\preprint{}

\title{Flat-Bands-Enabled Triplet Excitonic Insulator in a Di-atomic Kagome Lattice}

\author{Gurjyot Sethi\textsuperscript{1}}
\author{Yinong Zhou\textsuperscript{1}}%
\author{Linghan Zhu\textsuperscript{2}}
\author{Li Yang\textsuperscript{2,3}}
\author{Feng Liu\textsuperscript{1}}
\affiliation{%
 \textsuperscript{1}Department of Materials Science and Engineering, University of Utah, Salt Lake City, Utah 84112, USA\\
\textsuperscript{2}Department of Physics, Washington University in St. Louis, St. Louis, Missouri, 63130, USA\\
 \textsuperscript{3}Institute of Materials Science and Engineering, Washington University in St. Louis, St. Louis, Missouri, 63130, USA
}%

\date{\today}

\begin{abstract}
The excitonic insulator (EI) state is a strongly correlated many-body ground state, arising from an instability in the band structure towards exciton formation. We show that the flat valence and conduction bands of a semiconducting diatomic Kagome lattice, as exemplified in a superatomic graphene lattice, can possibly conspire to enable an interesting triplet EI state, based on density functional theory (DFT) calculations combined with many-body GW and Bethe-Salpeter Equation (BSE). Our results indicate that massive carriers in flat bands with highly localized electron and hole wavefunctions significantly reduce the screening and enhance the exchange interaction, leading to an unusually large triplet exciton binding energy ($\sim$1.1 eV) exceeding the GW band gap by $\sim$0.2 eV and a large singlet-triplet splitting of $\sim$0.4 eV. Our findings enrich once again the intriguing physics of flat bands and extend the scope of EI materials.
\end{abstract}

\maketitle


The discovery of excitonic insulator (EI) state has been a sought-after endeavor since it was first proposed by Kohn \cite{1,2} about fifty years ago. The EI phase is an exotic highly correlated electronic state that can be stabilized in narrow-gap semiconductors or semimetals \cite{1,2,3,4} via spontaneous formation of excitons below a critical temperature (T\textsubscript{c}). Originally Bardeen-Cooper-Schiefer (BCS) theory of superconductivity was used to model the EI state \cite{1,2} in the semi-metallic regime (negative band gap, E\textsubscript{g}), where a high carrier density makes the electron-hole (e-h) Coulomb's attraction strongly screened for a suppressed T\textsubscript{c} \cite{5}. On the other hand, for semiconductors, if the exciton binding energy (E\textsubscript{b}) exceeds E\textsubscript{g}, a spontaneous Bose-Einstein condensation (BEC) of excitons triggers the formation of EI state, and the coherence in bosonic wavefunctions leads to super transport \cite{4,7} and a weaker screening increases T\textsubscript{c} \cite{8}. The study of EI state should give deeper insight into highly correlated phenomena like superconductivity and BEC-BCS crossover \cite{9,10,11}, and a plethora of theoretical and experimental investigations have been made in an effort to realize this state \cite{8,12,13,14,15}. However, difficulty arises when trying to experimentally identify it since the excitons are neutral species whose “current” is not straightforwardly measurable. This demands investigation into other experimental signatures of EI \cite{8,13,14,15}.\par
The realization of an EI requires highly reduced screening to Coulomb's potential that leads to a higher E\textsubscript{b}. Low-dimensional materials tend to have reduced screening \cite{16,17} due to confinement effect. While two-dimensional (2D) semiconductors with a small E\textsubscript{g} may have lower E\textsubscript{b} because polarizability is inversely related to E\textsubscript{g} \cite{18}, dipole forbidden transitions near the band edges are shown to break this synergy and favor the formation of an intrinsic EI state, such as in 2D GaAs \cite{19} and Graphone \cite{20}. Another natural way of reducing screening is by increasing the e-h wavefunction overlap \cite{21,22,23}, which brings electrons and holes closer making them immune to the screening effect from surrounding charges.\par
Triplet EI state is especially appealing, because triplet excitons carry spin current so that a triplet EI with spin superfluidity can be experimentally observed by spin transport measurements \cite{24}. Triplet excitons are also attracting increasing attention in photovoltaics owing to their high radiative lifetime \cite{25,26}. Due to optical selection rule \cite{27}, the triplet excitons are dark but may be converted from singlets by intersystem crossing \cite{28,29}. A large singlet-triplet splitting ($\Delta$E\textsubscript{ST}) will favor such crossing process and increase the triplet concentration at finite temperatures \cite{30,31,32,33,34}. Hence, a large e-h wavefunction overlap is especially desirable for the triplet EI formation because it increases $\Delta$E\textsubscript{ST} \cite{31,32} by enhancing the e-h exchange interaction \cite{35}.\par
In this Letter, we demonstrate an intriguing FBs-enabled mechanism that can possibly lead to the formation of triplet EI mediated by massive carriers with greatly enhanced e-h wavefunction localization and overlap in a 2D diatomic (Yin-Yang) Kagome lattice \cite{36}, and further predict its realization in a real material made of superatomic graphene. Intrinsic to a topological flat band (FB) is its highly localized wavefunction in real space, underlined by a destructive interference of FB quantum states \cite{37}. When a unique band-structure configuration arises, with both a topological valence and conduction FB separated by a trivial gap, it also indicates an extremely high degree of e-h wavefunction overlap. Remarkably, the huge effective masses of carriers hosted in the two FBs greatly reduce the screening to increase exciton E\textsubscript{b}, favoring the formation of EI states in general; while the highly overlapping e-h wavefunction enhances the Coulomb's direct interaction on the one hand, to further reduce the screening, and the exchange interaction on the other hand, to increase $\Delta$E\textsubscript{ST}, favoring the formation of triplet EI states in specific. Using a superatomic graphene lattice as a prototypical example, we show the possibility of a FBs-enabled triplet EI state, based on DFT calculations combined with many-body GW and BSE. This is indicated by a triplet exciton E\textsubscript{b} ($\sim$1.1 eV) exceeding the GW gap (E\textsubscript{g} $\sim$0.9 eV) by $\sim$20\% and a high $\Delta$E\textsubscript{ST}  ($\sim$0.4 eV). FBs also provide an ideal platform for BEC and coherence \cite{38} since the macroscopic degeneracy of FBs can lead to spontaneous symmetry breaking which is central to the theory of BCS superconductivity \cite{39,40} and EI \cite{41,42}.\par
\begin{figure}[b]
\includegraphics[scale=0.38]{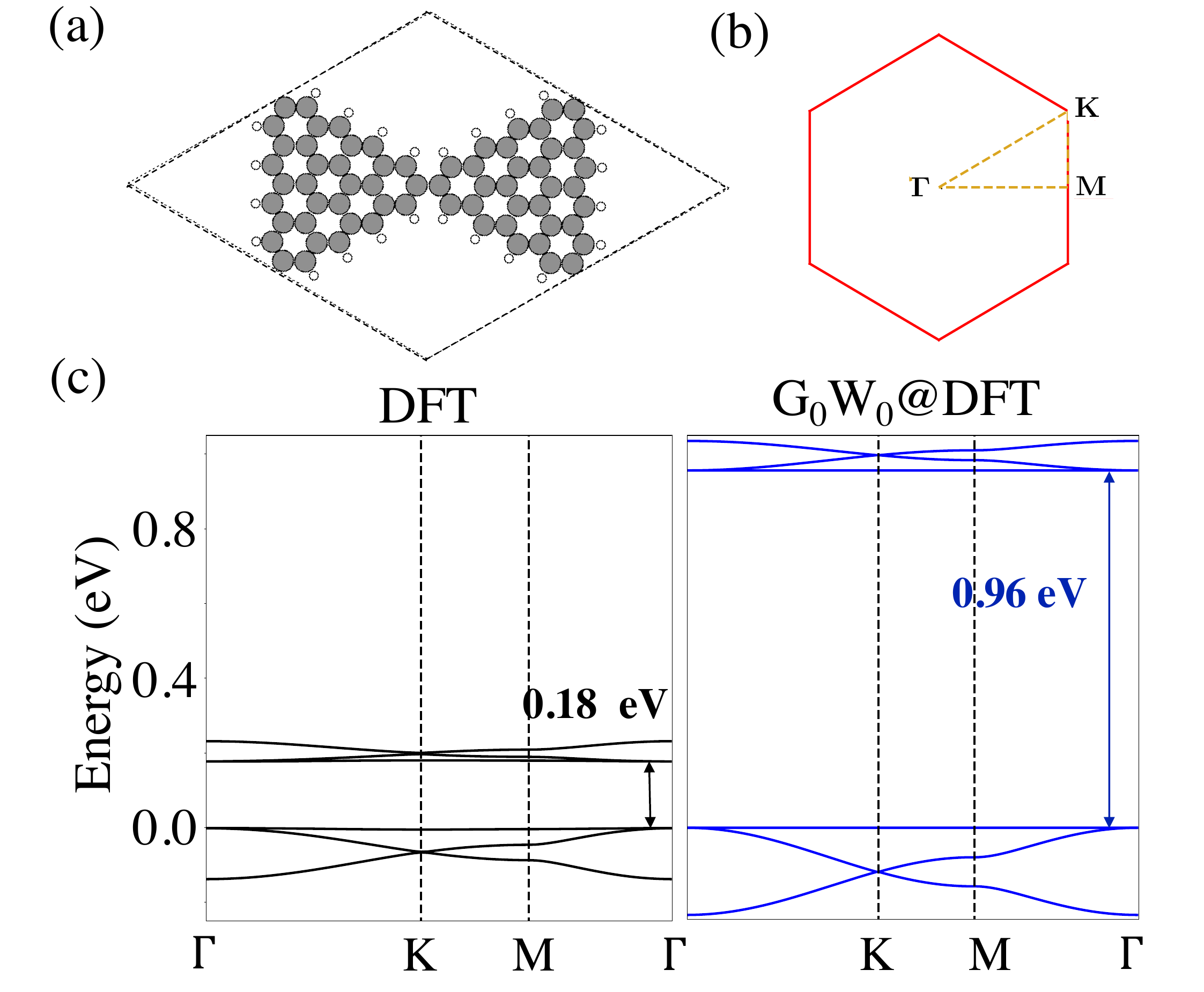}
\caption{\label{fig:fig1} (a) Unit cell of a 9 $\times$ 9 superatomic graphene lattice. Grey and white circles represent C and H atoms, respectively. (b) First Brillouin zone showing the high-symmetry reciprocal paths used for band diagrams. (c) Comparison of band structures and band gaps obtained within DFT (left panel) and a single-shot G\textsubscript{o}W\textsubscript{o} calculation (right).}
\end{figure}
Fig.~\ref{fig:fig1}(a) shows the structure of the superatomic graphene lattice. It consists of two 9$\times$9 grpahene flakes (structural motif) with an optimized lattice constant of 22.14 \AA. This peculiar structural motif enables the C atomic p\textsubscript{z} orbitals to hybridize into molecular sp\textsuperscript{2} orbitals, forming the so-called Yin-Yang Kagome bands \cite{36} in a hexagonal lattice, as shown in Fig.~\ref{fig:fig1}(c). Both the highest valence and the lowest conduction bands are perfectly flat and topologically nontrivial with opposite spin Chern numbers \cite{36}. The structure has a high thermodynamic stability with a bulk cohesive energy calculated as -6.78 eV per atom \cite{43}, similar to graphene nanoribbon \cite{44}. Excitingly, recent experimental advances \cite{45,46} in synthesizing nano-porous graphene suggest very high feasibility of making this lattice. These latest experiments employ a bottom-up approach to successfully make artificial nano-porous graphene lattices with precise control of pore size and shape, using designed molecular precursors. Accordingly, our theoretical study should stimulate such experimental efforts to make the proposed lattice by designing the desired superatomic graphene precursors. Also, other Ying-Yang Kagome lattices, such as the Kagome super-lattices formed in Moire pattern twisted graphene bilayers \cite{47}, which generated a lot of recent interest, can be generally explored. Here we focus on the exciton related properties of the 9$\times$9 superatomic graphene lattice.\par
\begin{figure}[b]
\includegraphics[scale=0.275]{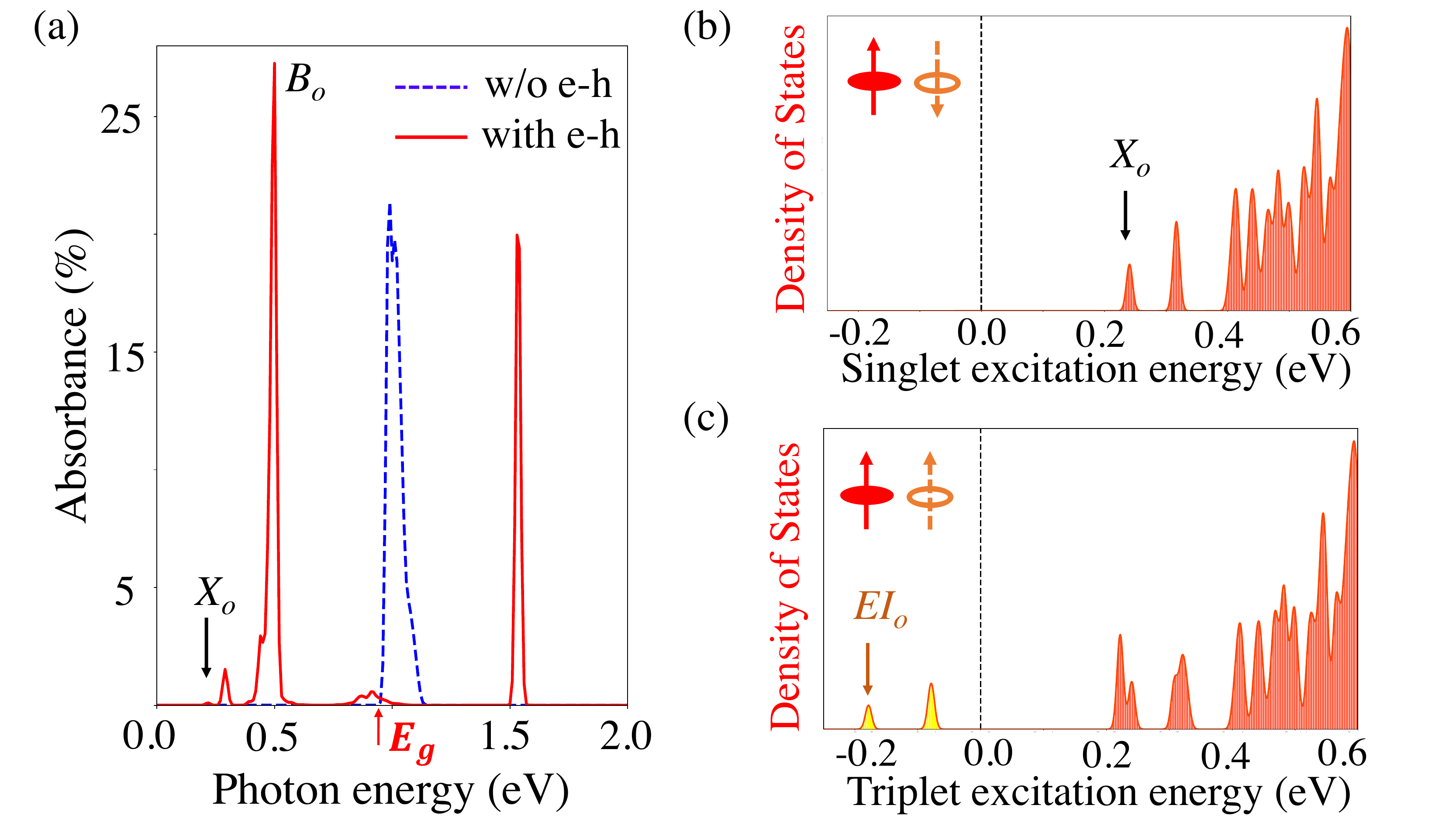}
\caption{\label{fig:fig2} (a) Optical absorbance for singlet excitons with a Gaussian peak broadening of 10 meV. \textit{X\textsubscript{o}} indicates the first bright peak. (b) Density of states for singlet excitons, and (c) Density of states for triplet excitons, noticing two peaks with negative formation energies (yellow-filled). \textit{EI\textsubscript{o}} indicates the first triplet exciton which is also the case of triplet EI. Insets in (b) and (c) represent spin-up electron (red arrow with filled circle) and spin-down hole (orange arrow with hollow circle) bound together forming an exciton.}
\end{figure}
First, its mean-field DFT E\textsubscript{g}$\sim$0.18 eV, as in a narrow-gap semiconductor, is significantly corrected to a GW E\textsubscript{g}$\sim$0.94 eV (Fig. ~\ref{fig:fig1}(c)). The optical spectra obtained by solving BSE is shown in Fig.~\ref{fig:fig2}(a), in comparison with that obtained within the independent particle approximation. The first peak in the BSE spectra corresponds to the first bright singlet exciton at 0.24 eV, marked by \textit{X\textsubscript{o}}. This can also be seen from the density of states (DOS) for singlet excitons in Fig.~\ref{fig:fig2}(b). The formation energy of \textit{X\textsubscript{o}} is very low compared to the quasiparticle gap of 0.94 eV, giving rise to a large E\textsubscript{b}  of 0.70 eV. In Fig.~\ref{fig:fig2}(c), we plot the DOS for triplet excitons which clearly shows the presence of excitons with negative formation energy, indicative of spontaneous formation of excitons. The E\textsubscript{b} of the lowest triplet exciton (0.94 eV + 0.17 eV = 1.11 eV) exceeds the GW gap by 0.17 eV, to signify a desired property for a strong triplet EI state, as marked by \textit{EI\textsubscript{o}} in Fig.~\ref{fig:fig2}(c). This possible triplet EI in a non-magnetic material is different from that recently studied in a ferromagnetic material where the excitation between spin non-degenerate bands was considered \cite{20}. One interesting feature is its huge $\Delta$E\textsubscript{ST} of 0.41 eV, making it easier to be detected by spin superfluidity experiment \cite{24}. In Table~\ref{tab:table1} we summarize the energies of the lowest singlet (\textit{X\textsubscript{o}}) and triplet (\textit{EI\textsubscript{o}}) states. The key result of a negative triplet exciton formation energy is carefully confirmed by convergence tests for GW-BSE calculations (see Table S1-S5 \cite{43}).\par
\begin{table*}
\caption{\label{tab:table1}%
Summary of excitonic energies for states \textit{X\textsubscript{o}} and \textit{EI\textsubscript{o}}. Last column denotes the dipole oscillator strength of excitons divided by the number of k-points.
}
\begin{ruledtabular}
\begin{tabular}{cccccc}
\textrm{Exciton}&
\textrm{\makecell{Mean field\\ gap (eV)}}&
\textrm{\makecell{GW band \\gap (eV)}}&
\textrm{\makecell{Excitation \\energy (eV)}}&
\textrm{\makecell{Binding \\energy (eV)}}&
\textrm{\makecell{Dipole Oscillator \\strength$/$N\textsubscript{k} ($\mu_s$)}}\\
\colrule
Singlet (\textit{X\textsubscript{o}}) & 0.18 & 0.96 & 0.21 & 0.75 & 0.024 au \\
Triplet (\textit{EI\textsubscript{o}}) & 0.18 & 0.96 & -0.21 & 1.17 & -\\
\end{tabular}
\end{ruledtabular}
\end{table*}
\begin{figure*}
\includegraphics[scale=0.52]{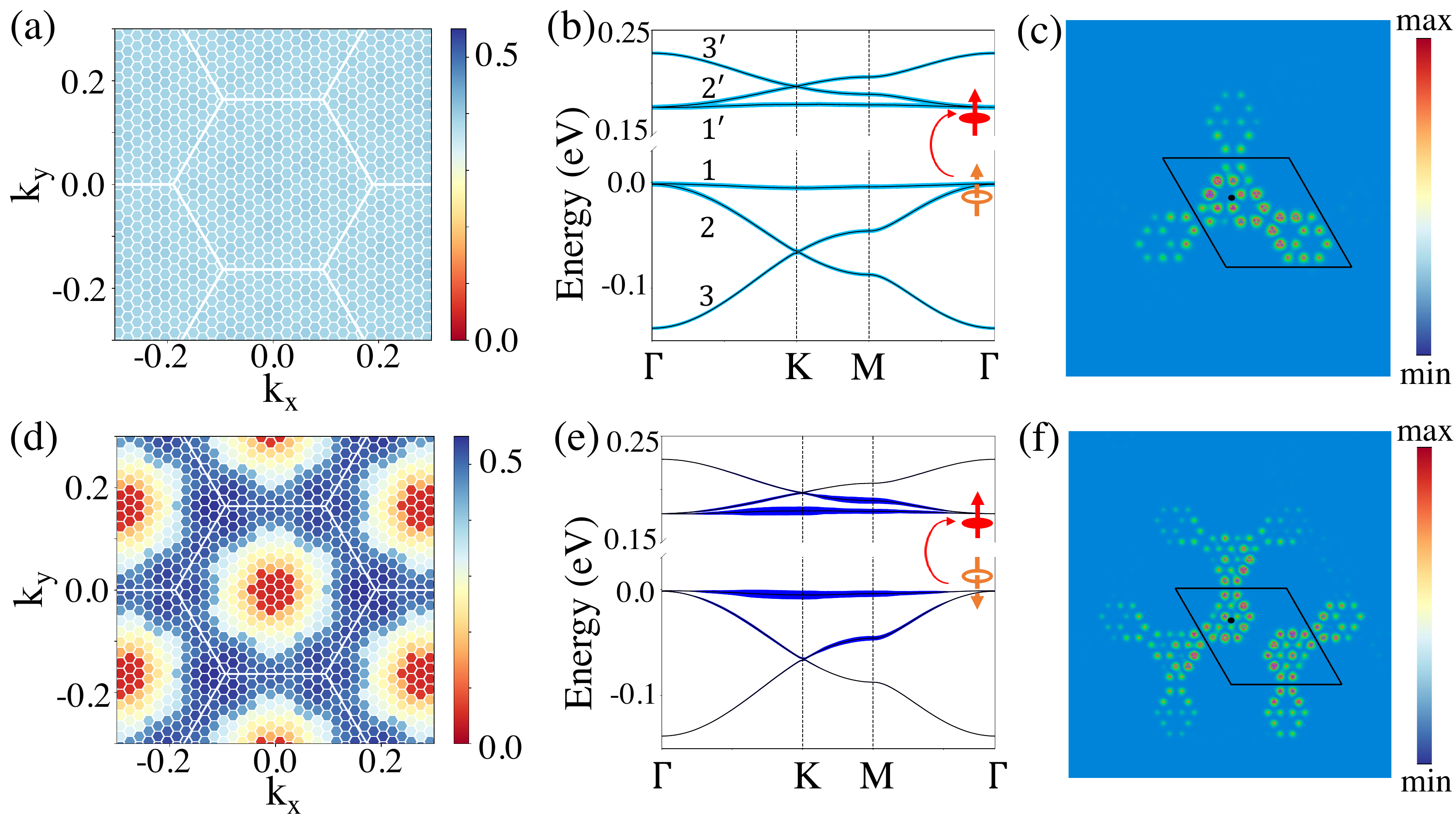}
\caption{\label{fig:fig3} Excitonic wavefunction analysis for the lowest triplet exciton (\textit{EI\textsubscript{o}}, Upper panel) and for the  lowest singlet exciton (\textit{X\textsubscript{o}}, lower panel): (a), (d) Reciprocal-space excitonic wavefunction distribution showing total contribution of all excitations at each reciprocal lattice point, (b), (e) Band excitation contributions indicated by spread of color on the bands,  (c), (f) Real space two-particle e-h wavefunction distribution with hole fixed at the black dot plotted over a 6$\times$6 supercell. Only a small segment of the supercell is shown here since the electron is highly localized around the hole. The distribution amplitude is zero everywhere else. The orange arrow with a circle denotes the spin of hole left behind after excitation and the red arrow denotes the spin of electron.}
\end{figure*}
Excitonic instability may occur in either a narrow-gap semiconductor or a semimetal \cite{48}. In the former, a gapped system as studied here, the critical condition for the existence of EI state is a negative exciton formation energy, i.e., if E\textsubscript{b} exceeds E\textsubscript{g}, the order parameter for BEC of excitons has a non-trivial solution at low temperatures. In a two-band model, the order parameter at T= 0 K is given by $\mathrm{\Delta_0=E_b\sqrt{(\frac{1}{2}-\frac{E_g}{2E_b})}}$ \cite{48}. As long as E\textsubscript{b}$>$E\textsubscript{g},  $\Delta_0$ is finite positive and a BEC-EI state emerges below T\textsubscript{c}. Our calculated E\textsubscript{b} is $\sim$20\% larger than E\textsubscript{g}, indicating a relatively high T\textsubscript{c}. For a system with parabolic valence and conduction bands, T\textsubscript{c} may be estimated from a $k.p$ effective-mass model \cite{20}. However, this model is not applicable to FBs with infinite effective mass, instead an extensive exact diagonalization approach is required to determine T\textsubscript{c}. Differently in a semimetal, the EI state occurs below T\textsubscript{c}\textsuperscript{'} via a metal (gapless) to insulator (gapped) transition, a manifestation of spontaneous symmetry breaking \cite{2,48}. It further involves a BCS-BEC crossover depending on the e-h coupling strength \cite{49}. One widely studied material for semimetal-to-EI transition is Ta\textsubscript{2}NiSe\textsubscript{5}, and a recent work \cite{42} shows that such transition may be generally triggered by a structural transition with breaking of lattice symmetries. Also, the BEC-BCS crossover has been studied in the context of non-equilibrium EI state \cite{51}.\par
In order to better understand the strikingly enhanced E\textsubscript{b}, suggestive of the formation of triplet EI ground state below T\textsubscript{c}, we plot the excitonic wavefunction in reciprocal space and band excitation contributions for \textit{EI\textsubscript{o}} in Fig.~\ref{fig:fig3}(a) and Fig.~\ref{fig:fig3}(b), respectively. One clearly sees that \textit{EI\textsubscript{o}} is composed of coherent excitations throughout the Brillouin zone (BZ). We also examined relative band contributions to the excitations (see Table S6 \cite{43}). The contribution from valence to conduction FB excitation is slightly higher than from other excitations. The nature of FB excitations inherently implies localized wavefunctions in real space as seen in Fig.~\ref{fig:fig3}(c), which shows the Fourier transform of excitonic wavefunction for \textit{EI\textsubscript{o}}, consistent with the broad distribution of k-point-resolved excitations (Fig. S1 \cite{43}). This provides additional evidence for a possible BEC-EI state, because the triplet exciton width ($\xi$) is much smaller than the lattice constant, implying a point-like boson, as in the BEC condensate \cite{51,52,53}. In comparison, for \textit{X\textsubscript{o}}, as shown in Fig.~\ref{fig:fig3}(d)-(e), excitation from valence to conduction FB contributes the most, largely centered around the K point. The triplet \textit{EI\textsubscript{o}} state has an even more localized wavefunction in real space (Fig.~\ref{fig:fig3}(c)) than the singlet \textit{X\textsubscript{o}} (Fig.~\ref{fig:fig3}(f)), because the former is excited throughout the BZ, i.e., the excitonic wavefunction is highly delocalized in reciprocal space.\par
The effective static dielectric constant obtained from our calculations is unusually low $\sim$1.02, indicating a highly reduced screening, which is a direct manifestation of FB wavefunctions as we show below. Usually dipole forbidden transitions near the band edges are favored for large E\textsubscript{b}, and hence the formation of EI state \cite{18,19,20}. For the Yin-Yang FBs, the inter-FB transitions are actually allowed by symmetry \cite{36} but the band flatness makes the dipole matrix element between them negligible. Considering a two-band model, the dipole matrix element is given by,
\begin{equation}
|<u_{c,k}|\nabla_k|u_{v,k}>|^2=\frac{\hbar^2}{2\mu}(E_g+\frac{\hbar^2k^2}{2\mu})^{-1}\label{eq:eqn1}
\end{equation}
where $\mu$ is the reduced mass under the effective mass approximation \cite{18}, which is very high here for both valence and conduction FBs making the above expression close to zero. A negligible dipole matrix element is directly verified from the absence of absorption peak at E\textsubscript{g} in the optical spectrum (Fig.~\ref{fig:fig2}(a)) obtained within the independent particle approximation (see also Fig. S4 \cite{43}). This also explains the very low absorbance for \textit{X\textsubscript{o}} (Fig.~\ref{fig:fig2}(a)) since the major contribution to this state is from FB excitations, and its non-zero portion of absorption is mostly contributed by weaker transitions that involve parabolic bands (e.g., 1$\rightarrow$2$'$ near M point) as shown in Fig.~\ref{fig:fig3}(e). Basically, only parabolic-band transitions lead to high optical absorption. The detailed contributions from individual band excitations are available in supplementary material \cite{43} (Table S1 and Fig. S2), and the band resolved contributions to the brightest exciton, marked in Fig.~\ref{fig:fig2}(a) by \textit{B\textsubscript{o}}, is available in Fig. S3 \cite{43}. Since 2D polarizability is proportional to the dipole matrix element divided by the gap \cite{18}, the presence of FBs as both the highest occupied valence and the lowest unoccupied conduction bands inherently reduces the screening significantly.\par
\begin{figure}[b]
\includegraphics[scale=0.38]{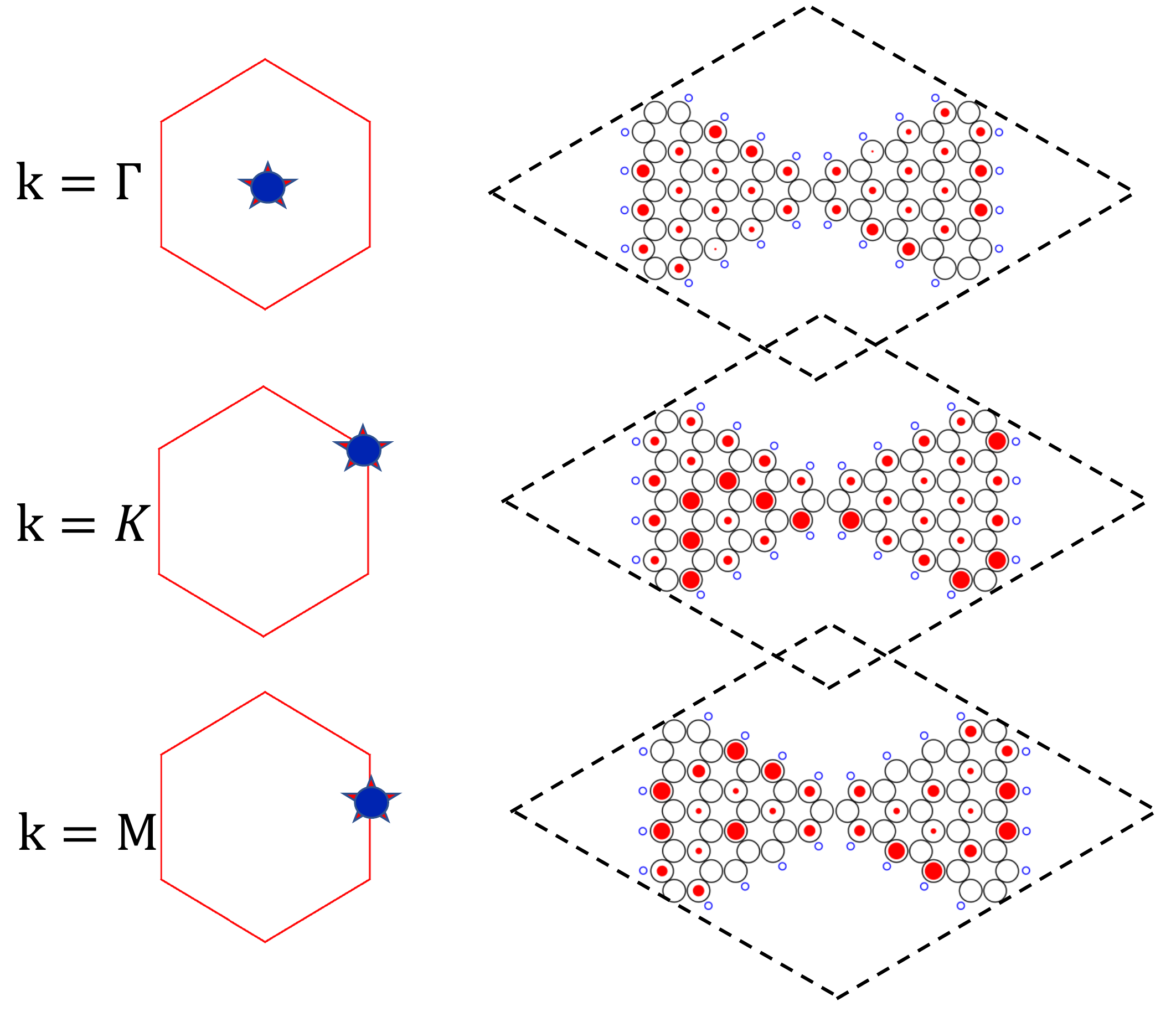}
\caption{\label{fig:fig4} Conduction and valence FBs wavefunction overlap for zero momentum singlet exciton (Q = 0). Right: C and H atoms are indicated by large and small circles respectively. The contributions are only from the C p\textsubscript{z} orbitals. The overlapped weights of these contributions to flat valence and conduction bands wavefunctions are indicated by the size of red fills on the C atoms. Left: k-points in BZ for which overlaps are calculated.}
\end{figure}

Normally, reduced screening and confinement effects in low dimensions are known to extrinsically increase the e-h wavefunction overlap as shown in hetero-nanostructures \cite{30} and carbon nanotubes \cite{31}. Interestingly here both electron and hole wavefunctions exhibit a form of destructive quantum interference originated from the FB topology \cite{37}, which gives rise to their distinguished localized states in real space with huge overlap. Fig.~\ref{fig:fig4} shows the relative overlap between the two FB wavefunctions at high-symmetry k-points projected over atomic orbitals of C and H. Such huge overlap leads to a much higher energy for singlet compared to triplet excitons as the exchange interaction is absent in the latter. This represents a unique \textit{intrinsic} FB originated increase of direct and exchange energy, leading to a $\Delta$E\textsubscript{ST} of $\sim$0.4 eV, much larger than the typical values in bulk (a few meVs) \cite{30} and low-dimensional semiconductors (upto $\sim$0.2 eV) \cite{31,32,33,34}. Thus, both a huge e-h wavefunction overlap and a highly reduced screening, as induced by the FBs, are the major factors leading to a large E\textsubscript{b}, and an enhanced $\Delta$E\textsubscript{ST}, favorable for a triplet EI. More singlet exciton properties are given in SM \cite{43}. \par
For comparison, it might be interesting to construct a supermolecule consisting of two superatomic graphene flakes and study its excitonic properties, for which a Frenkel type of localized exciton is expected. Also, in the condensed state of Van der Waals molecular crystals, excitons with relatively large E\textsubscript{b} and almost linear excitonic dispersion have been previously shown \cite{54}. In contrast, for the covalently bonded framework we study here, the highly localized and overlapping exciton wavefunctions, as shown in Fig.~\ref{fig:fig3}(c) and Fig.~\ref{fig:fig3}(f), are enabled by FBs, having a very different physical origin. This is because the FB is a Bloch state; it is encoded with nontrivial topology, arising from destructive interference of lattice hopping that leads to compact plaquette states of both electrons and holes in real space \cite{36}. As such, the excitons display an unusual constant dispersion (see also the discussion for singlet excitons in SM \cite{43}).\par
In conclusion, we reveal a unique topological FB-originated mechanism for the possible formation of a spin triplet EI, making a significant forward step for the discovery of EIs through spin transport measurement. In a 9 $\times$ 9 superatomic graphene lattice, a triplet exciton E\textsubscript{b} is predicted to exceed the band gap by $\sim$0.2 eV. The FBs, generally existing in a Yin-Yang Kagome lattice, weaken intrinsically the screened e-h interaction by an “infinite” effective mass of carriers and a complete overlap of e-h wavefunctions, and the latter also increases the exchange energy of singlet exciton leading to a huge $\Delta$E\textsubscript{ST}. In general, defects are likely present in experimental samples. However, if the defect density is kept low, there is usually no significant change in the screening and excitonic properties, as shown in transition metal dichalcogenides \cite{55}. Spin-orbit coupling (SOC) may have interesting consequences in our proposal; however, since only C and H atoms are present here, it is negligible and not considered. Other Ying-Yang Kagome lattice materials with large SOC are interesting topics of future studies, especially in considering the related phenomena of excited quantum anomalous and spin Hall effect \cite{36}. Furthermore, fractional population of two FBs may lead to exotic fractional EI state \cite{56}.\par
G.S, Y.Z. and F.L. acknowledge financial support from US Department of Energy-Basic Energy Sciences (Grant No. DE-FG02- 04ER46148). L.Z. and L.Y. are supported by the Air Force Office of Scientific Research (AFOSR) grant No. FA9550-20-1-0255 and the National Science Foundation (NSF) CAREER grant No. DMR-1455346. The calculations were done on the CHPC at the University of Utah and DOE-NERSC.


\section*{Supplementary material}

\section{Computational Details}
We carried out density-functional-theory (DFT) calculations using Perdew-Burke-Ernzerhof (PBE) exchange-correlation functional \cite{1s} as implemented in Quantum Espresso code \cite{2s}. The plane-wave energy cutoff is set at 50 Ry based on norm-conserving pseudopotentials with the semicore electrons included that amounts to a system with 144 valence electrons per unit cell. We choose a 6 $\times$ 6 $\times$ 1 k-grid sampling in the reciprocal space. A vacuum thickness of $\sim$44 $\mathrm{\AA}$  is used to avoid spurious interactions between adjacent layers. Atomic positions were fully relaxed until residual forces were less than 0.1 meV/$\mathrm{\AA}$. GW-BSE simulations are performed by using the BerkeleyGW code \cite{3s} including the slab Coulomb truncation. The quasiparticle energies and band gaps are calculated within the single shot G\textsubscript{o}W\textsubscript{o} approximation using the general plasmon pole model \cite{4s}. The optical absorption spectra and excitonic effects are calculated by solving the BSE \cite{5s}. We use a coarse k-point grid of 5 $\times$ 5 $\times$ 1 for calculating the e-h interaction kernel and a fine k-point grid of 16 $\times$ 16 $\times$ 1 with a total of 720 bands for converged bandgap, excitonic states and optical absorption spectra. For optical absorption spectra, 3 valence and 3 conduction bands were used and only the incident light polarized parallel along the plane is considered due to the depolarization effect along the out-of-plane direction. The BSE is solved for zero center-of-mass momentum of excitons.

\section{Convergence of GW-BSE}
In order to obtain meaningful physical insight from GW and GW+BSE calculations, convergence tests for the GW band gap and exciton binding energies are performed with respect to the following parameter settings: 
\begin{itemize}
	\item Number of empty conduction bands and energy cutoff for dielectric matrix (epsilon$\_$cutoff) used in the ‘epsilon’ code
	\item Energy cutoff for bare Coulomb interaction (bare\_coulomb\_cutoff) used in ‘sigma’ code to calculate the GW band gap
	\item K-point mesh
	\item Number of conduction and valence bands used in ‘absorption’ code to calculate excitonic properties
\end{itemize}\par
We first test for the number of empty bands used for calculating the dielectric matrix with a k-point mesh of 5$\times$5, epsilon\_cutoff of 6 Ry and bare\_cloumb\_cutoff of 50 Ry as shown in Table~\ref{tab:tables1}. For all the subsequent tests this is fixed at 720. We then test for the epsilon\_cutoff with k-point mesh fixed at 5$\times$5 and bare Coulomb’s cutoff fixed at 50 Ry. The results are presented in Table~\ref{tab:tables2}. Based on this convergence test we choose 6 Ry as the optimized epsilon\_cutoff for further tests. Similarly, we tested for bare\_coulomb\_cutoff (Table~\ref{tab:tables3}) and choose 40 Ry as the optimized value.\par
The above optimized values are then used to test for k-point mesh convergence (Table~\ref{tab:tables4}). With a k-pint mesh of 4$\times$4 the calculations already start to approach the converged GW band gap and binding energies for both singlet and triplet excitons. Finally, we optimize the number of conduction and valence bands used for calculating excitonic properties as shown in Table~\ref{tab:tables5}.\par
Following are the optimized values of parameters used in this work:
\begin{itemize}
	\item Number of empty conduction bands in ‘epsilon’ – 720
	\item Energy cutoff for dielectric matrix– 6 Ry
	\item Energy cutoff for bare Coulomb’s interaction – 40 Ry
	\item K-point mesh - 5×5
	\item 3 conduction and 3 valence bands for BSE
\end{itemize}
\section{Material Stability}
In order to estimate the thermodynamic stability of the 9$\times$9 superatomic graphene lattice with molecular formula C\textsubscript{66}H\textsubscript{24} considered here, we estimate its bulk cohesive energy as follows,
\begin{equation}
E_{cohesive}=\frac{E_{C_{66}H_{24}}-66E_C-24E_H}{90}
\end{equation}
where $E_{C_66H_24}$ is the energy of the unit cell,  $E_C$ and $E_H$ are the energies of individual carbon and hydrogen atoms, respectively. The calculated value of -6.78 eV is comparable to the cohesive energies for graphene, benzene and graphene nanoribbons \cite{6s} indicating a high thermodynamic stability of the material considered in this work. 
\section{Excitonic wavefunctions}
The solution of BSE Hamiltonian in reciprocal space gives the excitonic binding energy and excitonic wavefunctions, which one can visualize to understand the contribution of excitations to an excitonic state at each k-point in the Brillouin Zone (BZ). The BSE equation under Tamm-Dancoff approximation is \cite{3s},
\begin{eqnarray}
(E_{ck}^{QP}-E_{vk}^{QP})A_{vck}^S+\sum_{v'c'k'}<vck|K_{eh}|v'c'k'>=E_S A_{vck}^S\;,
K_{eh}=K_{eh}^d+K_{eh}^x
\end{eqnarray} 
where $A_{vck}^S$ is the excitonic wavefunction (in the quasiparticle state representation), $E_S$ is the excitation energy (also referred to as the formation energy) of the exciton in state S, and $K_{eh}$ is the electron-hole interaction kernel which is the sum of direct and exchange interaction kernel. In Fig. 3(a) and 3(d) we plot the sum,
\begin{equation}
A_k^S=\sum_{vc}A_{vck}^S
\end{equation}
over the whole BZ for the lowest singlet and triplet states respectively. One can also plot $A_{vck}^S$ for each excitation v$\rightarrow$c at each point in BZ for any excitonic state S. Here we plot $A_{vck}^S$ for S=\textit{EI\textsubscript{o}} (Fig.~\ref{fig:figs1}),  S=\textit{X\textsubscript{o}} (Fig.~\ref{fig:figs2}), and S=\textit{B\textsubscript{o} }(Fig.~\ref{fig:figs3}), specifically for individual excitation contribution to these states at each k point. \par
$W_{tot}$ is defined as the sum of excitonic wavefunction weights ($A_{vck}^S$ ) for an excitation over the whole BZ,
\begin{equation}
W_{tot}=\sum_{k\subset BZ}A_{vck}^S
\end{equation}\par
These excitonic weights for \textit{EI\textsubscript{o}} and \textit{X\textsubscript{o}} are listed in Table~\ref{tab:tables6} and Table~\ref{tab:tables7} respectively.
\section{Optical spectrum within the independent particle approximation}
The dipole matrix element for the flat band transitions is negligible so that there is no peak in the optical spectrum obtained within the independent particle approximation, as shown in Fig. 2(a), at the photon energy that equals to band gap. To further ascertain this, we calculate the absorbance for various transitions in the tight-binding framework of Yin-Yang Kagome lattice, and plot over the first BZ. The interband optical absorbance for the light polarized along x-direction is given by \cite{7s},
\begin{equation}
A_{v,c}(k)=\int d\omega \frac{\alpha_o}{\hbar \omega}|<\psi_c(k)|\frac{\partial H(k)}{\partial k_x}|\psi_v(k)>|^2\times\frac{\eta}{\pi[(E_c(k)-E_v(k)-\hbar\omega)^2+\eta^2]}
\end{equation}
where $|\psi_n(k)>$ is the valence ($n=v$) or conduction ($n=c$) band wavefunction, $\eta$ is the full width at half maximum of the Lorentzian, $\alpha_o=e^2/4\pi \epsilon_o c \hbar$, and $\omega$ is the frequency of light. The flat-band transitions have negligible contribution to the absorbance as shown in Fig.~\ref{fig:figs4}.
\section{Real-space electron-hole pair correlation function}
The e-h pair correlation function of triplet excitons in real space is highly localized as shown in Fig. 3(c). We note that the real-space distribution of exciton wavefunction, as the way calculated by arbitrarily choosing a "point" hole position, does not keep the lattice symmetry in general and changes with the hole position \cite{8s}. But its degree of localization, i.e., the range of e-h pair correlation function, is well represented and insensitive to the hole position. As shown in Fig.~\ref{fig:figs5} for triplet exciton using two different hole positions, the electrons are highly localized around the hole within the area of one unit-cell. In contrast, for singlet exciton as shown in Fig.~\ref{fig:figs6}, the electrons are less localized around the hole, spreading beyond the area of one unit-cell. We also note that the exciton wavefunction distribution in the reciprocal space keeps the lattice symmetry as shown in Fig. 3(a) and Fig. 3(d).
\section{Singlet exciton and Electron-hole wavefunction overlap}
Here we show that the singlet excitons, although not forming an EI state, have a much-enhanced intrinsic lifetime arising from the FB excitations. The radiative lifetime for an zero-momentum exciton in state S with an excitation energy E\textsubscript{s}(0) and a zero excitonic momentum can be estimated for 2D systems as \cite{9s}, 
\begin{equation}
\tau_o=\frac{A_{uc}\hbar^2c}{8\pi e^2E_s(0)\mu_s^2}\label{eq:eqn2}
\end{equation}
where $\mu_s$ is the exciton dipole oscillator strength per number of k-points ($N_k$) and $A_{uc}$ is the area of unit cell. Using the BSE calculated $\mu_s$, we obtain $\tau_o$ for \textit{X\textsubscript{o}} to be 80.9 ps, which is very high compared to excitons in transition metal dichalcogenides \cite{9s} and single-walled carbon nanotubes \cite{10s}. The finite temperature-averaged radiative lifetime depends on excitonic dispersion.\par
\begin{figure}[H]
\includegraphics[scale=0.5]{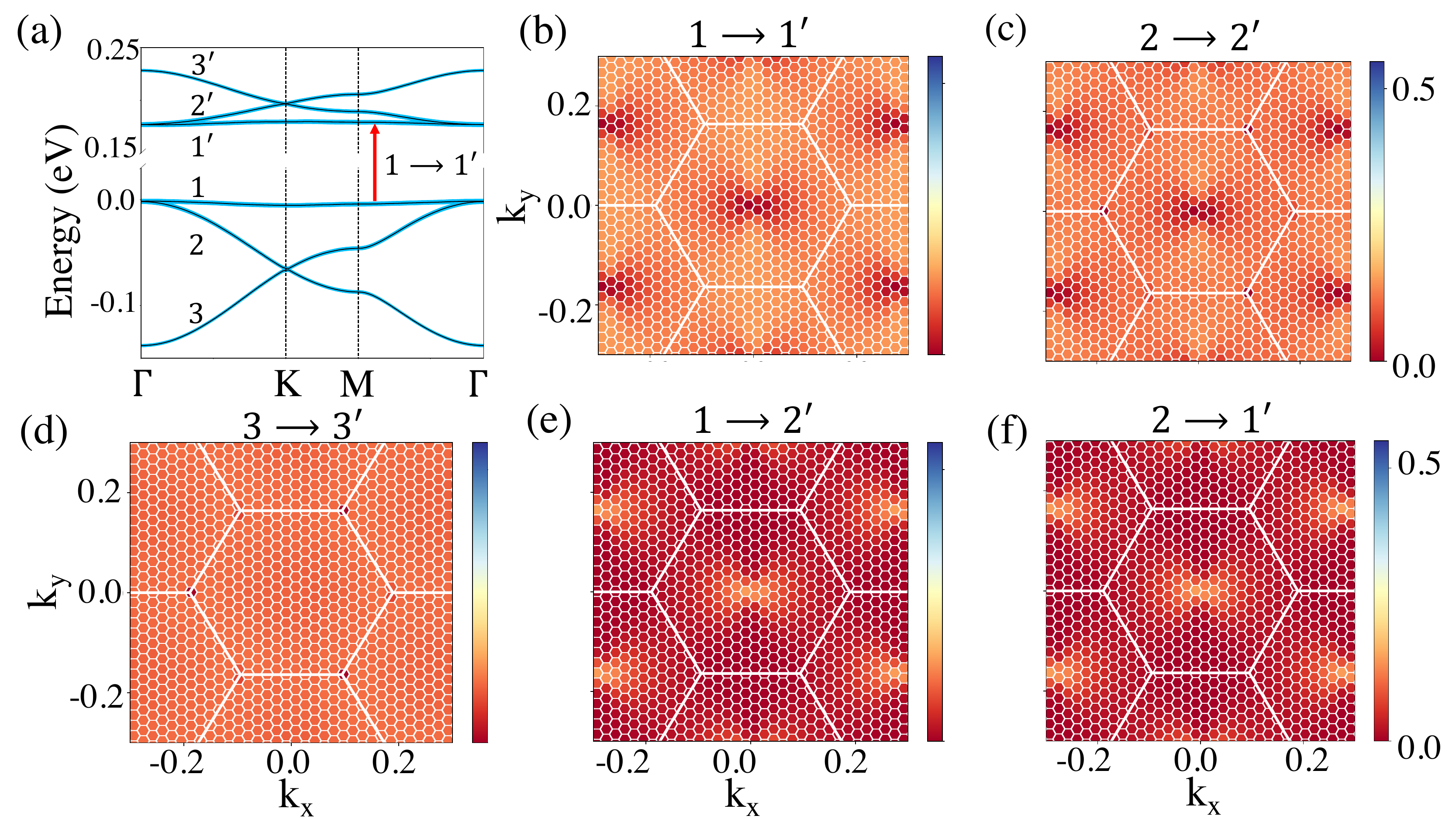}
\caption{\label{fig:figs1} Individual excitation contribution to \textit{EI\textsubscript{o}}. The band structure in (a) has the valence and conduction bands numbered, according to which we label the excitations. The excitation 1$\rightarrow$1$'$ is shown for reference. In (b), (c), (d), (e) and (f) we plot the contribution of excitations corresponding to 1$\rightarrow$1$'$, 2$\rightarrow$2$'$, 3$\rightarrow$3$'$, 1$\rightarrow$2$'$, and 2$\rightarrow$1$'$ respectively. There is a complete population inversion between the lower Kagome bands to upper enantiomorphic Kagome bands. At the $\Gamma$ point in the first BZ, the bands 1 (1$'$) and 2 (2$'$) are degenerate for conduction  (valence) bands, which causes a slight decrease in the contribution of excitations 1$\rightarrow$1$'$ and 2$\rightarrow$2$'$ at $\Gamma$ point. This is compensated by the excitations 1$\rightarrow$2$'$ and 2$\rightarrow$1$'$ which contribute only at $\Gamma$ point. Contributions from other excitations are negligible and hence not shown here.}
\end{figure}
\begin{figure}[H]
\includegraphics[scale=0.5]{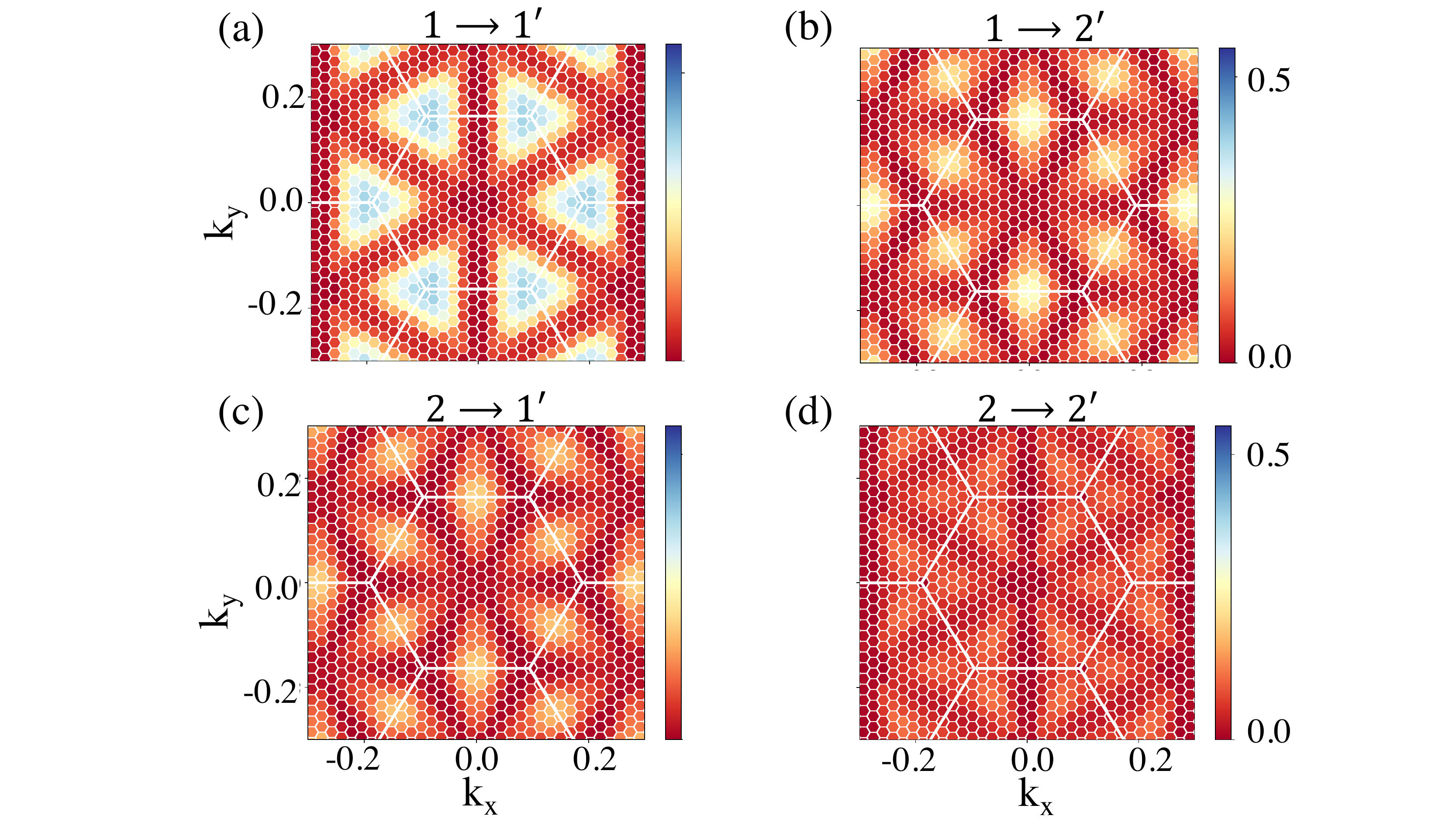}
\caption{\label{fig:figs2} Individual excitation contribution to \textit{X\textsubscript{o}}. The bands are numbered the same way as in Fig.~\ref{fig:figs1}. Excitation 1$\rightarrow$1$'$ contributes the most to this excitonic state. Contributions from other excitations are negligible and hence not shown here.}
\end{figure}
\begin{figure}[H]
\includegraphics[scale=0.5]{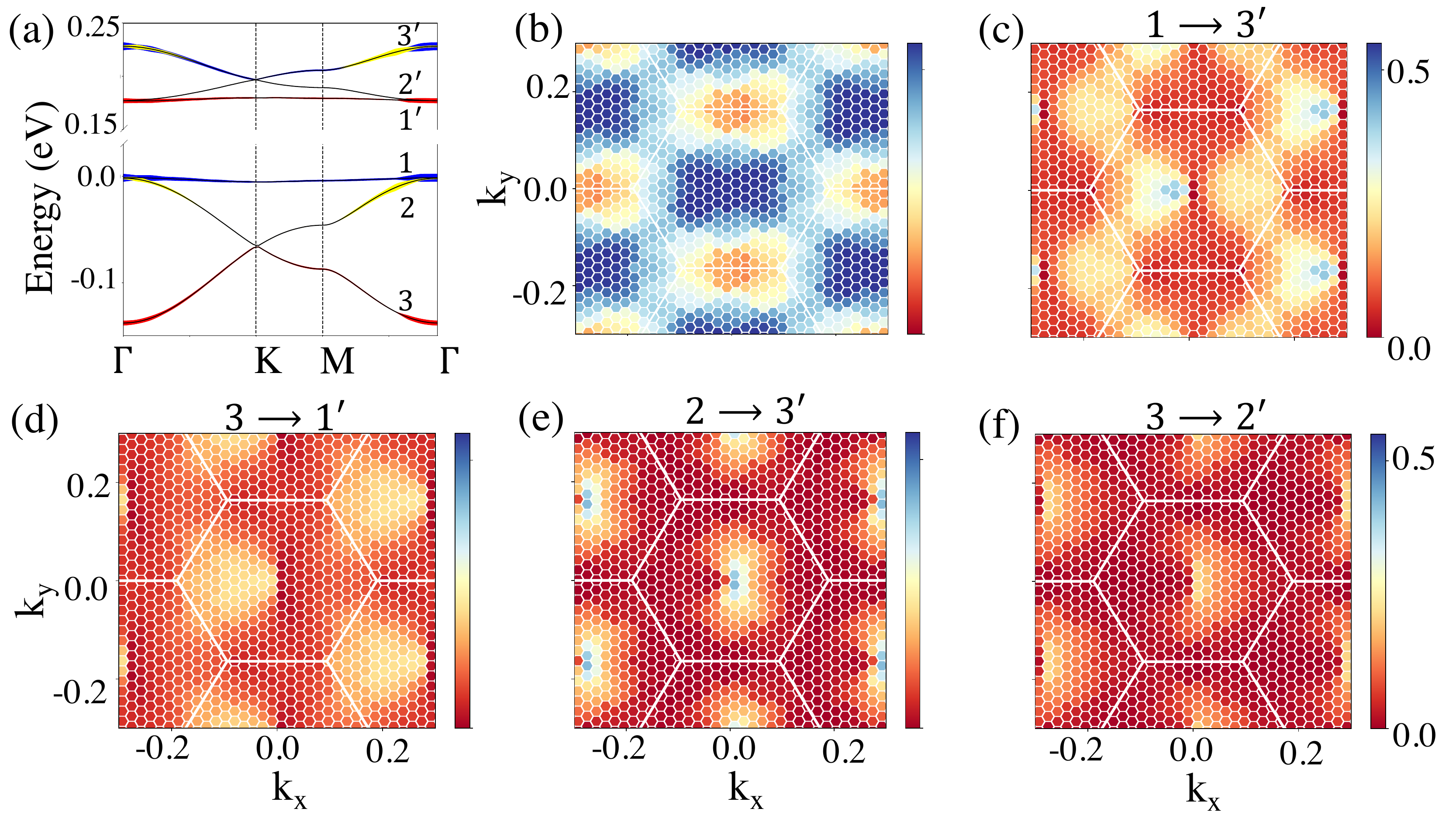}
\caption{\label{fig:figs3} Individual excitation contributions to \textit{B\textsubscript{o}}. (a) Band excitation contributions. The colors indicate the individual excitations (blue - 1$\rightarrow$3$'$, red - 3$\rightarrow$1$'$, and yellow - 2$\rightarrow$3$'$). (b) Total contribution of all excitations at each k-point, (c)-(f) k-space resolved contribution for each excitation. There is no contribution from inter-flat-bands transition. Contribution from parabolic bands makes this state the brightest. Contributions from other excitations are negligible and hence not shown here.}
\end{figure}

Towards understanding the excitonic exchange and direct interaction as one varies the excitonic center-of-mass momentum, we plot the overlapped weights of atomic orbitals that contribute to the flat valence and conduction bands wavefunctions at various k-points in the BZ. The wavefunctions projected over atomic orbitals only have contributions from the C p\textsubscript{z} orbitals, which also point to the hybridization of these orbitals into forming an effective sp\textsuperscript{2} molecular orbital in hexagonal lattice. As shown in Fig. 4, the distribution is highly spread out at each k-point and the wavefunctions near K point have the highest overlap between conduction and valence FBs, while the overlap is negligible for $\Gamma$ point owing to their degeneracy with parabolic bands. This is consistent with the excitonic wavefunction distribution plotted in Fig. 3(d) and Fig. 3(e) with the maximum and minimum weight located at K and $\Gamma$ point, respectively, for singlet \textit{X\textsubscript{o}}, which depend on both direct and exchange interactions. The e-h wavefunction overlaps for triplet \textit{EI\textsubscript{o}} also agree with the distribution plotted in Fig. ~\ref{fig:figs1} which is only determined by the direct Coulomb interaction.\par
These plots can help reveal the singlet excitonic dispersion relations. The direct ($M_{c,c}^G (k+Q,q)\times M_{v,v}^G (k+Q,q)$) and exchange ($M_{c,v}^G (k+Q,q)$) coulombic matrix elements for non-zero center-of-mass momentum of excitons, which go into the BSE Hamiltonian \cite{5s}, depend on the following integrals,
\begin{equation}
M_{c,c}^G(k+Q,q)=\int dx \psi_{c,k+Q}^*(x) e^{-i(q+G).r}\psi_{c,k+Q+q}(x)
\end{equation}
\begin{equation}
M_{v,v}^G(k+Q,q)=\int dx \psi_{v,k}^*(x) e^{-i(q+G).r}\psi_{v,k+q}(x)
\end{equation}
\begin{equation}
M_{c,v}^G(k+Q,q)=\int dx \psi_{c,k+Q}^*(x) e^{-i(q+G).r}\psi_{v,k+q}(x)
\end{equation}
Here $\psi_{n,k} (x)$ is the Bloch’s wavefunction for valence (n=v) or conduction band (n=c), and Q is the center-of-mass momentum of the exciton, i.e., the electron is excited from the k point in the valence flat band to the k+Q point in the conduction flat band. We therefore plot the overlapped wavefunctions between valence flat band at k and conduction flat band k+Q, for high-symmetry k-points in the BZ.\par
In Fig.~\ref{fig:figs7} and Fig.~\ref{fig:figs8}, the schematic in the left panel shows the pair of k-points in BZ for which the wavefunctions are overlapped. The k-point considered for valence wavefunction (v,k) is marked by a red star, while the corresponding k+Q point for conduction wavefunction (c,k+Q) is marked by a blue circle. The overlapped weights of these contributions to valence and conduction band wavefunctions are indicated by the size of red fills on the C atoms in the right panel. As can be seen from Fig. 4, Fig.~\ref{fig:figs7} and Fig.~\ref{fig:figs8}, the overlaps are huge and do not change as one changes Q from $\Gamma$ to K. From Q=K to Q=M, there is change in overlap for individual  k-points but the total remains the same.\par
This indicates that the singlet FB exciton is almost dispersion-less, i.e., it has a very large effective mass. In order to consider the effect of finite temperature on radiative lifetime one can average the decay rate over the range of momentum accessible at temperature T. This statistical average is given by \cite{9s},
\begin{equation}
<\tau_{eff}>=\tau_o\frac{3}{4}(\frac{k_B T}{\Delta_s})\label{eq:eqn3}
\end{equation}
where $\Delta_s$ is the difference between the exciton binding energy for Q=0 and Q=Q\textsubscript{o}, the maximum momentum a decaying exciton can have \cite{10s}. Because $\Delta_s$ is negligible here, one expects the already enhanced singlet exciton lifetime to be much higher at finite temperatures by allowing excitons with Q$>$Q\textsubscript{o} to be thermally activated easily.



\section{Supplementary tables}
\begin{table}[H]
\caption{\label{tab:tables1}%
Convergence test for \# of empty bands used in calculating the dielectric matrix with k-point mesh fixed at  5$\times$5, epsilon\_ctuoff fixed at 6 Ry and bare\_coulomb\_cutoff fixed at 50 Ry. The excitonic properties are calculated for 3 valence and 6 conduction bands.
}
\begin{ruledtabular}
\begin{tabular}{cccc}
\textrm{\# empty bands in `epsilon'}&
\textrm{GW E\textsubscript{g} (eV)}&
\textrm{Triplet E\textsubscript{b} (eV)}&
\textrm{Singlet E\textsubscript{b} (eV)}\\
\colrule
500 & 0.96 & 1.14 & 0.71 \\
720 & 0.94 & 1.11 & 0.70\\
\end{tabular}
\end{ruledtabular}
\end{table}

\begin{table}[H]
\caption{\label{tab:tables2}%
Convergence test for energy cutoff used in dielectric matrix with k-point mesh fixed at  5$\times$5, and bare\_coulomb\_cutoff fixed at 50 Ry. The excitonic properties are calculated for 3 valence and 6 conduction bands.
}
\begin{ruledtabular}
\begin{tabular}{cccc}
\textrm{epsilon\_cutoff}&
\textrm{GW E\textsubscript{g} (eV)}&
\textrm{Triplet E\textsubscript{b} (eV)}&
\textrm{Singlet E\textsubscript{b} (eV)}\\
\colrule
5 & 0.93 & 1.10 & 0.69\\
6 & 0.94 & 1.11 & 0.70\\
8 & 0.94 & 1.11 & 0.70\\
\end{tabular}
\end{ruledtabular}
\end{table}

\begin{table}[H]
\caption{\label{tab:tables3}%
Convergence test for energy cutoff used in bare Coulomb’s interaction with k-point mesh fixed at  5$\times$5, and epsilon\_cutoff fixed at 6 Ry. The excitonic properties are calculated for 3 valence and 6 conduction bands.
}
\begin{ruledtabular}
\begin{tabular}{cccc}
\textrm{bare\_coulomb\_cutoff}&
\textrm{GW E\textsubscript{g} (eV)}&
\textrm{Triplet E\textsubscript{b} (eV)}&
\textrm{Singlet E\textsubscript{b} (eV)}\\
\colrule
40 & 0.94 & 1.11 & 0.70\\
50 & 0.94 & 1.11 & 0.70\\
\end{tabular}
\end{ruledtabular}
\end{table}

\begin{table}[H]
\caption{\label{tab:tables4}%
Convergence test for k-point mesh with epsilon\_cutoff fixed at 6 Ry, and bare\_coulomb\_cutoff fixed at 40 Ry. The excitonic properties are calculated for 3 valence and 6 conduction bands.
}
\begin{ruledtabular}
\begin{tabular}{cccc}
\textrm{k-point mesh}&
\textrm{GW E\textsubscript{g} (eV)}&
\textrm{Triplet E\textsubscript{b} (eV)}&
\textrm{Singlet E\textsubscript{b} (eV)}\\
\colrule
3$\times$3 & 1.05 & 1.36 & 0.92\\
4$\times$4 & 0.96 & 1.17 & 0.75\\
5$\times$5 & 0.94 & 1.11 & 0.70\\
\end{tabular}
\end{ruledtabular}
\end{table}

\begin{table}[H]
\caption{\label{tab:tables5}%
Convergence test for number of conduction (c) and valence (v) bands used in BSE with k-point mesh fixed at 5$\times$5,  epsilon\_cutoff fixed at 6 Ry, and bare\_coulomb\_cutoff fixed at 40 Ry.
}
\begin{ruledtabular}
\begin{tabular}{ccc}
\textrm{Bands in BSE}&
\textrm{Triplet E\textsubscript{b} (eV)}&
\textrm{Singlet E\textsubscript{b} (eV)}\\
\colrule
c = 3 \; v = 3 & 1.11 & 0.70\\
c = 6 \; v = 5 & 1.11 & 0.70\\
\end{tabular}
\end{ruledtabular}
\end{table}

\begin{table}[H]
\caption{\label{tab:tables6}%
Weights of individual band excitation contributions to the triplet excitonic state \textit{EI\textsubscript{o}}. The valence bands are numbered from top-down while the conduction bands are numbered from bottom-up as also shown in Fig.~\ref{fig:figs1}(a). Contributions from other excitations are negligible and hence omitted here.
}
\begin{ruledtabular}
\begin{tabular}{ccc}
\textrm{Valence band}&
\textrm{Conduction band}&
\textrm{$W_{tot}$}\\
\colrule
1 & 1' & 0.304 \\
2 & 2' & 0.271\\
3 & 3' & 0.270\\
1 & 2' & 0.068\\
2 & 1' & 0.067\\
\end{tabular}
\end{ruledtabular}
\end{table}

\begin{table}[H]
\caption{\label{tab:tables7}%
Weights of individual band excitation contributions to the singlet excitonic state \textit{X\textsubscript{o}} similar to values given in Table~ref{tab:tables6} for \textit{EI\textsubscript{o}}}
\begin{ruledtabular}
\begin{tabular}{ccc}
\textrm{Valence band}&
\textrm{Conduction band}&
\textrm{$W_{tot}$}\\
\colrule
1 & 1' & 0.359 \\
1 & 2' & 0.200\\
2 & 1' & 0.163\\
2 & 2' & 0.137\\
1 & 3' & 0.074\\
3 & 1' & 0.033\\
\end{tabular}
\end{ruledtabular}
\end{table}
\section{Supplementary figures}
\centering

\begin{center}
\begin{figure}[H]
\includegraphics[scale=0.5]{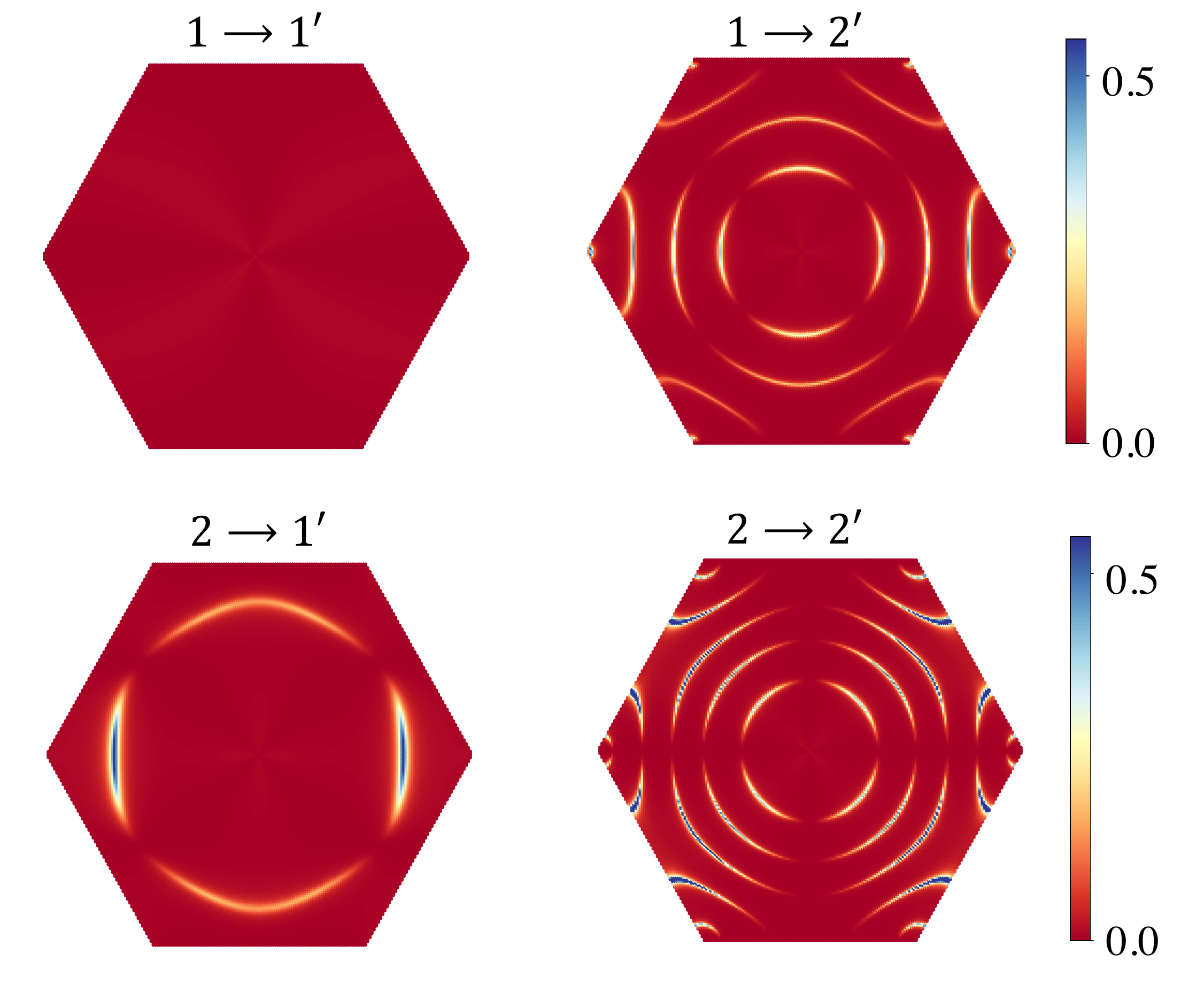}
\caption{\label{fig:figs4} Interband absorbance for first few transitions plotted over the first BZ for transitions labeled according to numbers shown in Fig.~\ref{fig:figs1}(a). The dipole matrix element for flat band transitions ($1\rightarrow1'$) is negligible.}
\end{figure}

\begin{figure}[H]
\includegraphics[scale=0.5]{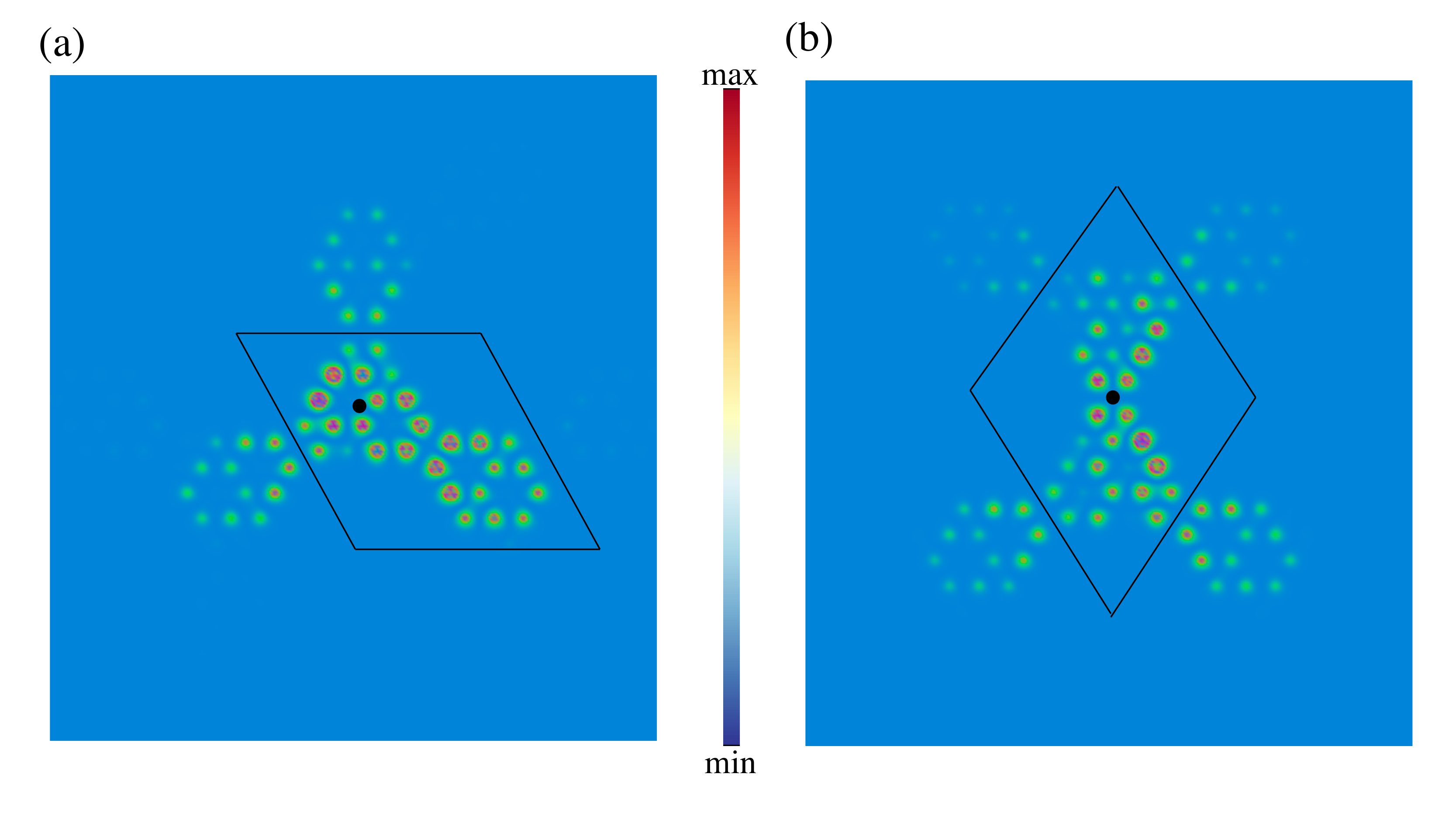}
\caption{\label{fig:figs5} Real space electron-hole pair correlation function for triplet excitonic state \textit{EI\textsubscript{o}} with hole (denoted by black dot) placed at center of graphene flake ((a)) and in between two flakes ((b)).}
\end{figure}

\begin{figure}[H]
\includegraphics[scale=0.5]{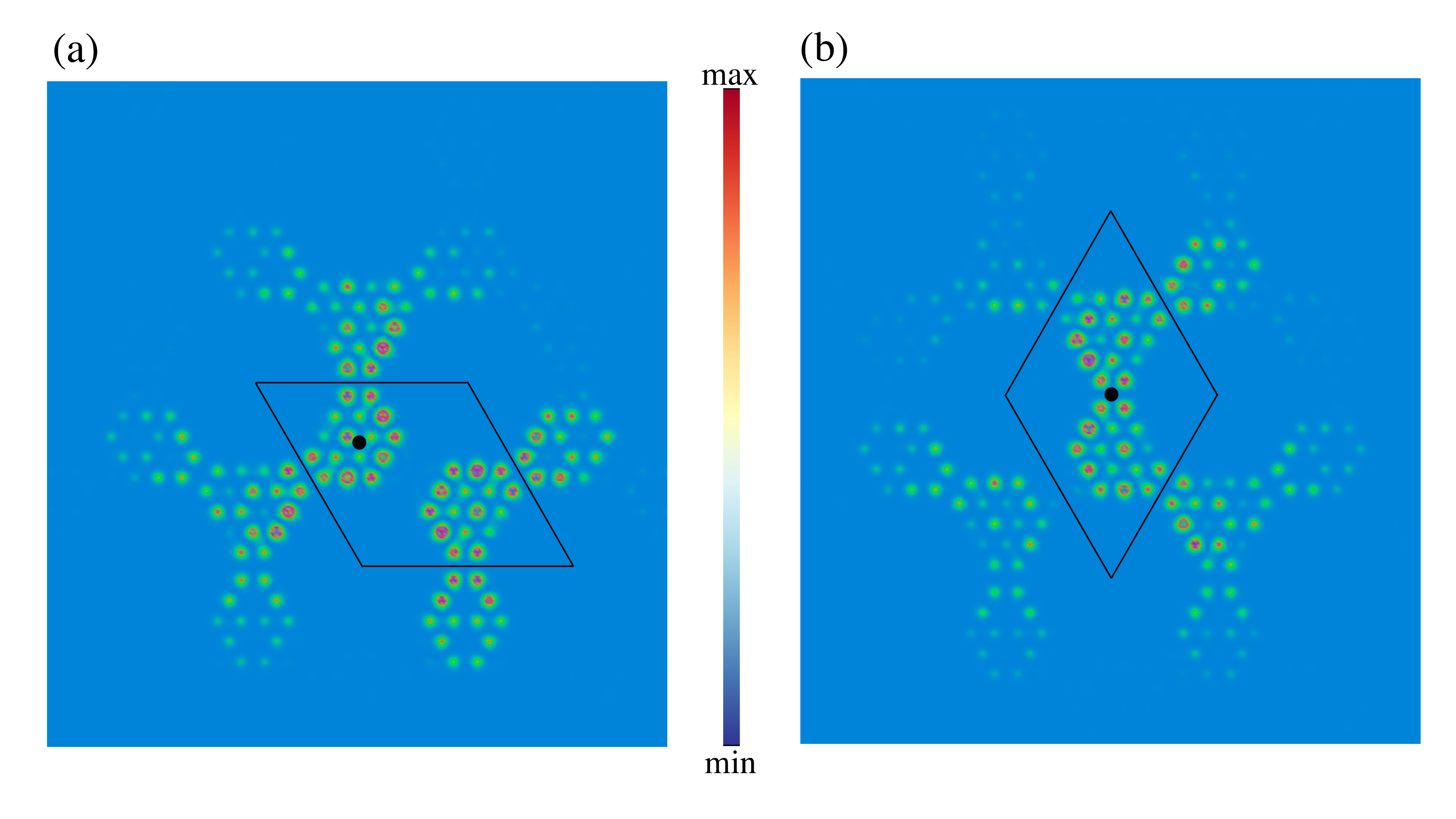}
\caption{\label{fig:figs6} Real space electron-hole pair correlation function for singlet excitonic state \textit{X\textsubscript{o}} with hole (denoted by black dot) placed at center of graphene flake ((a)) and in between two flakes ((b)).}
\end{figure}

\begin{figure}[H]
\includegraphics[scale=0.5]{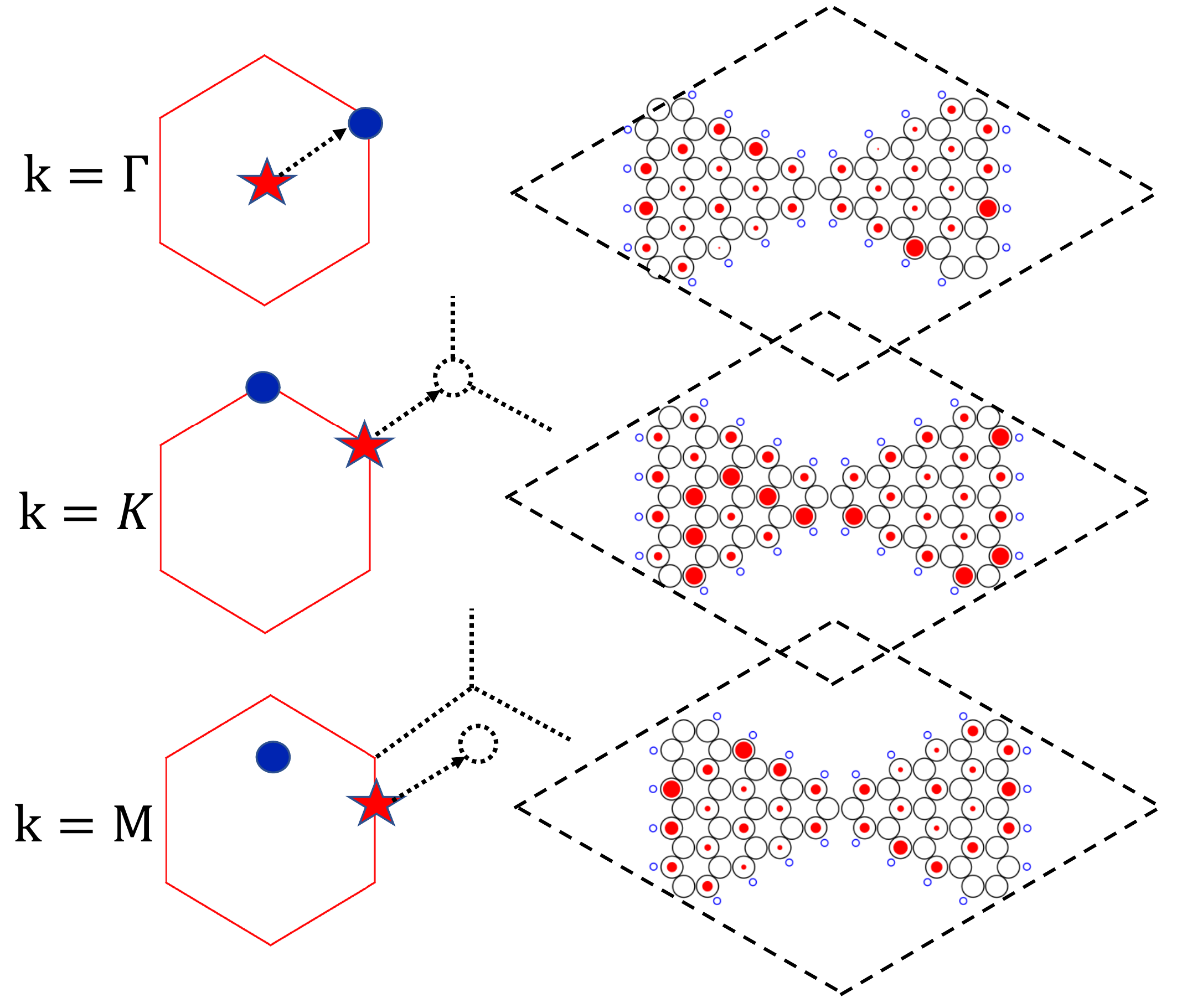}
\caption{\label{fig:figs7} The flat valence (hole) and conduction (electron) band wavefunction overlaps for Q=K. The dotted lines in the left panel show the equivale k-point (dashed circle) in the adjacent BZ at which the conduction band wavefunction is plotted.}
\end{figure}

\begin{figure}[H]
\includegraphics[scale=0.5]{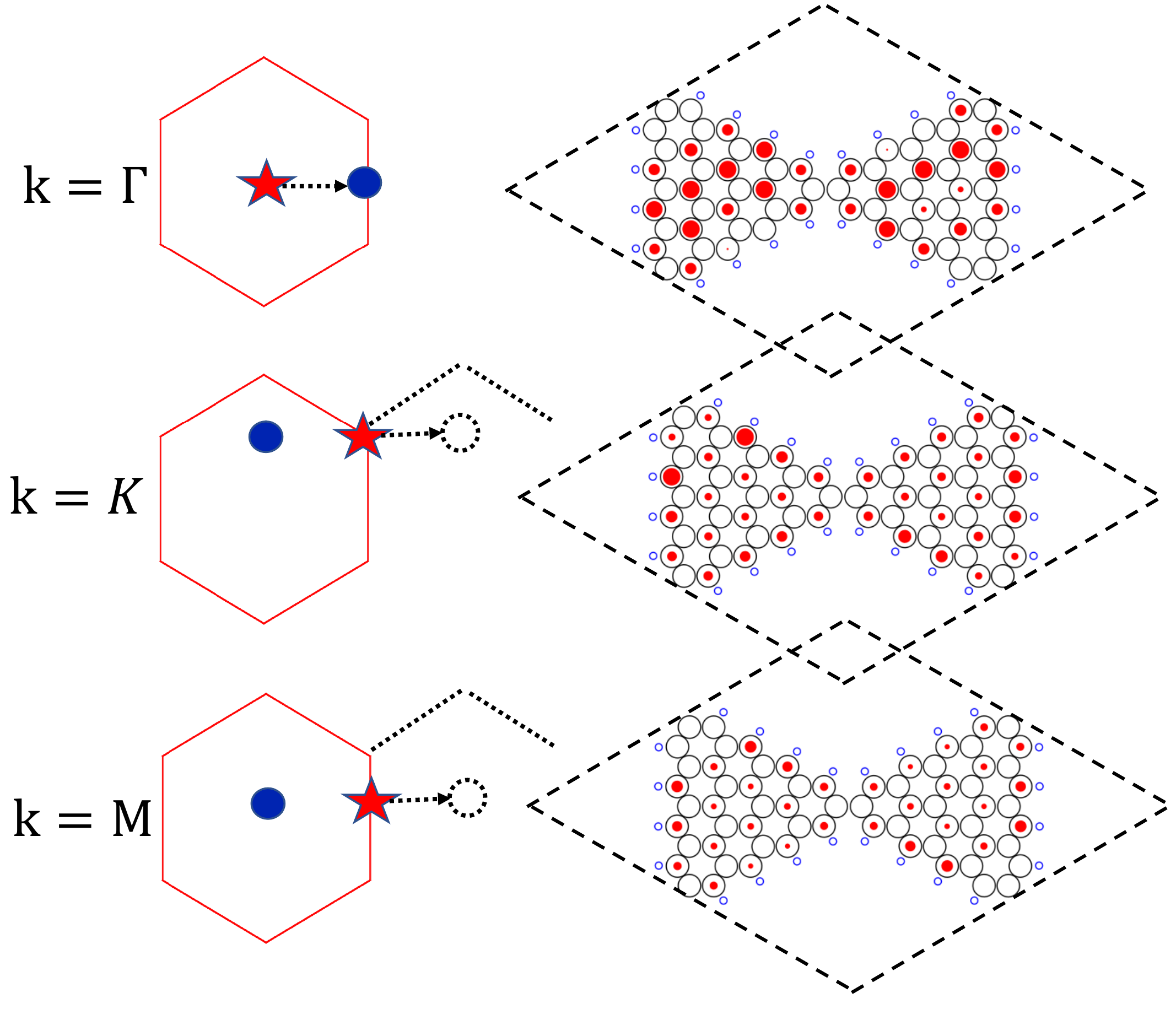}
\caption{\label{fig:figs8} The flat valence (hole) and conduction (electron) band wavefunction overlaps for Q=M. The dotted lines in the left panel show the equivalent k-point (dashed circle) in the adjacent BZ at which the conduction band wavefunction is plotted.}
\end{figure}
\end{center}

\bibliography{manuscript}

\end{document}